 \documentstyle[preprint,prl,aps,psfig]{revtex}         

\newcounter{FAQ}
\begin{document}
\draft

\title{The two-dimensional Anderson model of localization with random
  hopping}

\author{Andrzej Eilmes\\
Department of Computational Methods in Chemistry,
Jagiellonian University, 30-060 Krak\'{o}w, Poland,\\
Telephone: +48 12 336377 ext.\ 213, FAX: +48 12 340515
}
\author{Rudolf A.\ R\"{o}mer\\
Institut f\"{u}r Physik,
Technische Universit\"{a}t Chemnitz,
D-09107 Chemnitz, Germany,\\
Telephone: +49 371 531 3149, FAX: +49 371 531 3151
}
\author{Michael Schreiber\\
Institut f\"{u}r Physik,
Technische Universit\"{a}t Chemnitz,
D-09107 Chemnitz, Germany,\\
Telephone: +49 371 531 3142, FAX: +49 371 531 3143
}

\date{Version: June 18, 1997; printed \today}
\maketitle

\begin{abstract}
  We examine the localization properties of the 2D Anderson
  Hamiltonian with off-diagonal disorder. Investigating the behavior
  of the participation numbers of eigenstates as well as studying
  their multifractal properties, we find states in the center of the
  band which show {\em critical}\/ behavior up to the system size $N=
  200 \times 200$ considered.  This result is confirmed by an
  independent analysis of the localization lengths in quasi-1D strips
  with the help of the transfer-matrix method. Adding a very small
  additional onsite potential disorder, the critical states become
  localized.
\end{abstract}

\tighten

%
%

\section{Introduction}
\label{sec-intro}

Nearly forty years have passed since Anderson's \cite{anderson} first
suggestion of a disorder-induced metal-insulator transition (MIT), and
yet the localization problem remains at the center of much interest.

For non-interacting electrons, a highly successful approach was put
forward in 1979 by Abrahams {\em et al.} \cite{go4}. This ``scaling
hypothesis of localization'' suggests that an MIT exists for
non-interacting electrons in three dimensions (3D) at zero magnetic
field $B$ and in the absence of spin-orbit coupling.  Much further
work has subsequently supported these scaling arguments both
analytically and numerically \cite{rama,kramer}. In 1D and 2D, the
same hypothesis shows that there are no extended states and thus no
MIT.  However, since 2 is the lower critical dimension of the
localization problem \cite{wegener}, the 2D case is in a sense
``close'' to 3D: states are only marginally localized for weak
disorder and a small magnetic field or spin-orbit coupling can lead to
the existence of extended states and thus an MIT. Consequently, the
localization lengths of a 2D system with potential disorder can be
quite large \cite{pichard,mckinnon} so that in numerical approaches
one can always find a localization-delocalization transition when
decreasing either system size for fixed disorder or disorder for fixed
system size \cite{mueller}.

The role played by many-particle interactions is much less understood
\cite{belitz}.  Even for disordered quantum many-body systems in 1D,
no entirely consistent picture exists \cite{gs}.
Thus recent experimental results \cite{kravch}, which indicate the
existence of an MIT in certain 2D electron gases at $B=0$, are a
challenge to our current understanding.  In the samples considered,
the Coulomb interaction is estimated to be much larger than the Fermi
energy \cite{kravch} and so the observed MIT may be due to an
interaction-driven enhancement of the conductivity.  A recent
reevaluation \cite{abra97} of the principles of scaling theory shows
that these experimental results do in fact not violate general scaling
principles. However, it is not yet clear that this transition does
indeed correspond to an MIT since other recent arguments \cite{ist}
suggest that the transition might be understood as an
insulator-superconductor transition.

Most numerical approaches to the localization problem use the standard
tight-binding Anderson Hamiltonian with onsite potential disorder.
Characteristics of the electronic eigenstates are then investigated by
studies of participation numbers \cite{part} obtained by exact
diagonalization, multifractal properties \cite{mfa,mfarec}, level
statistics \cite{level} and many others. Especially fruitful is the
transfer-matrix method (TMM) \cite{mckinnon} which allows a direct
computation of the localization lengths and further validates the
scaling hypothesis by a numerical proof of the existence of a
one-parameter scaling function.

In the present work, we have reconsidered a variant of the Anderson
model in which also the nearest-neighbor hopping elements are allowed
to be randomly distributed. Prior to the advent of the scaling
hypothesis, Thouless-type arguments showed the possibility of much
larger localization lengths in such a 2D system \cite{LT75} as
compared to the case with potential disorder only. Thus, motivated in
part by the above mentioned experimentally observed transition in 2D
systems at $B=0$, this model provides a good starting point for a
search of 2D states which, perhaps, need not be localized.
Another motivation is provided by the 2D random magnetic flux model
(RFM) \cite{2drfm,batsch}, in which the hopping elements are chosen to
be of unit modulus but with a random phase representing a random
magnetic field penetrating the 2D plane. Although much effort has been
dedicated towards the RFM, no definite picture exists and results
range from a complete absence of diffusion to the prediction of
extended states near the band center \cite{2drfm}. Our random hopping
model may be viewed as a RFM with phase fixed at zero but random
modulus.

In the present paper, we will present a comprehensive numerical study
of the 2D Anderson model with random hopping. In section
\ref{sec-model} we introduce the model and notation. In order to get a
first insight into the differences {\em and} the similarities of the
random hopping and the potential disorder case, we look at the
eigenstates and their Fourier transforms in section \ref{sec-wavefcn}.
We calculate the density of states (DOS) in section \ref{sec-dos} and
show an unusual feature in the band center $E=0$. In section
\ref{sec-locprop} we then study participation numbers and multifractal
properties, respectively.  A scaling analysis of the participation
numbers suggests that the states at $E=0$ for system sizes up to $N=
200 \times 200$ behave similar to critical states at the MIT in the 3D
Anderson model. We confirm this result by the TMM together with the
one-parameter finite-size-scaling (FSS) analysis \cite{mckinnon} in
section \ref{sec-fssloclen}: the states at $E=0$ show critical
behavior up to a strip width $M=180$. However, already a very small
additional onsite potential disorder destroys the criticality.  We
summarize and conclude in section \ref{sec-concl}.

%
%

\section{The Model}
\label{sec-model}

The 2D Anderson Hamiltonian is given as
\begin{equation}
\label{eq-hamil}
  H = \sum_i^N \epsilon_i |i \rangle \langle i| +
      \sum_{i \neq j}^{N} t_{ij} |i \rangle \langle j|.
\end{equation}
The sites $i=(n,m)$ form a regular square lattice of size $N= L \times
L$ and, unless stated otherwise, we will always use periodic boundary
conditions. The onsite potential energies $\epsilon_i$ are taken to be
randomly distributed in the interval $[-W/2,W/2]$.  The transfer
integrals $t_{ij}$ are restricted to nearest-neighbors and {\em chosen
  to be randomly distributed in the interval} $[ c - w/2, c + w/2 ]$.
Thus $c$ represents the center and $w$ the width of the off-diagonal
disorder distribution.  We set the energy scale by keeping $w=1$
fixed, except for the cases of pure diagonal disorder where the
hopping elements are constant ($w=0$ and $c=1$). For $c \rightarrow
\infty$, the off-diagonal disorder width $w$ is negligible compared to
its mean, and we get the usual Anderson model; when additionally $W$
remains finite for $c\rightarrow\infty$, the system becomes ordered.
On the other hand, for $c \leq 0.5$, individual hopping elements may
be zero and transport will be hindered more strongly.  This will give
a more pronounced tendency towards localization.

We note that for the case of purely off-diagonal disorder ($W=0$) we
have an {\em exact} particle-hole symmetry in the band such that for
any eigenstate with energy $E_j>0$, there is also an eigenstate with
energy $E_{j'} = - E_j$. In the usual Anderson model with $W>0$ and
$w=0$, this exact symmetry is only recovered in the limit of infinite
system size $N\rightarrow\infty$.

Our numerical approach to the present model is based on (i) an exact
diagonalization of the respective secular matrices by means of the
Lanczos algorithm \cite{lanczos}, and (ii) a recursion form for the
Schr\"odinger equation corresponding to the Hamiltonian
(\ref{eq-hamil}) which provides the starting point for the TMM of
section \ref{sec-fssloclen}.

%
%

\section{Looking at the wave functions}
\label{sec-wavefcn}


\subsection{Probability density in real space}
\label{subsec-pars}

\begin{figure}
  \centerline{\psfig{figure=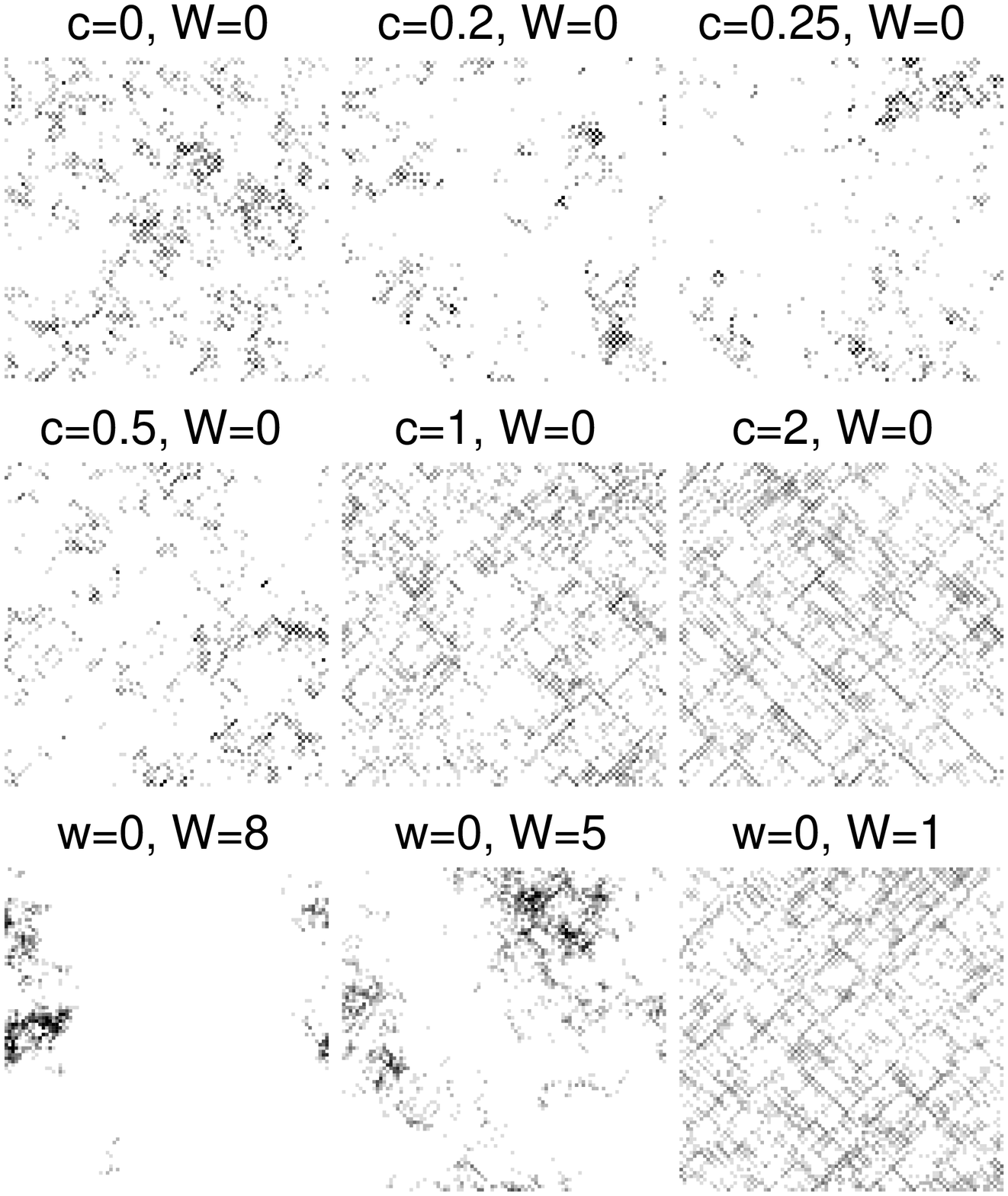,width=5.5in,height=6.4in}}
  \caption{
    Probability density $|\phi_j|^2$ of the eigenstate $j$ closest to
    the band center for various off-diagonal and diagonal disorders
    and system size $L=96$. Different gray levels ($i=0,1,\ldots,6$)
    distinguish whether $|\phi_j(n,m)|^2 > 2^i/L^2$.}
\label{fig-xphi}
\end{figure}

Let us start our investigation of the Hamiltonian (\ref{eq-hamil}) by
simply looking at some typical eigenstates $\phi_j(n,m)$ obtained by
exact diagonalization.
For small off-diagonal disorder $c > 0.5$, the ordered system is only
slightly perturbed and we expect the weakest localization of the wave
function to occur at the band center $E=0$ just as for purely diagonal
disorder $W$. In Fig.\ \ref{fig-xphi}, we show the spatial dependence
of the probability density of the wave function at $E=0$ for various
values of $c$. For $c > 0.5$, the probability density is rather
homogeneously distributed over all $N$ sites.  For comparison, we also
include 3 examples of an analogous plot for purely diagonal disorder
($w=0$), showing a similarly homogeneous distribution for small $W$.
E.g.\ the probability density plot at $c=2$ $(W=0)$ is very similar to
the plot for $W=1$ ($w=0$).
With decreasing $c \leq 0.5$ the wavefunctions become concentrated in
certain areas, indicating a tendency towards localization.  Moreover,
differences between diagonal and off-diagonal disorder become
noticeable: systems with purely off-diagonal disorder exhibit large
site-to-site probability density fluctuations resulting in
characteristic chess board patterns, whereas in the systems with
diagonal disorder separate areas of large probability appear. We also
see from Fig.\ \ref{fig-xphi} that $c = 0$ does not seem to correspond
to the strongest localization. Rather, the strongest ``curdling''
\cite{mfa} of a state occurs at $c \approx 0.25$.
For purely diagonal disorder, it is well-known that the localization
is strongest for states with energies close to the band edges.  In
agreement, we have found, but refrain from showing corresponding plots
here, that with increasing energy towards the band edges the patterns
of probability density for purely off-diagonal disorder tend to be
more localized and become thus again similar to those for diagonal
disorder.


\subsection{Probability density in Fourier space}
\label{subsec-pafs}

\begin{figure}
  \centerline{\psfig{figure=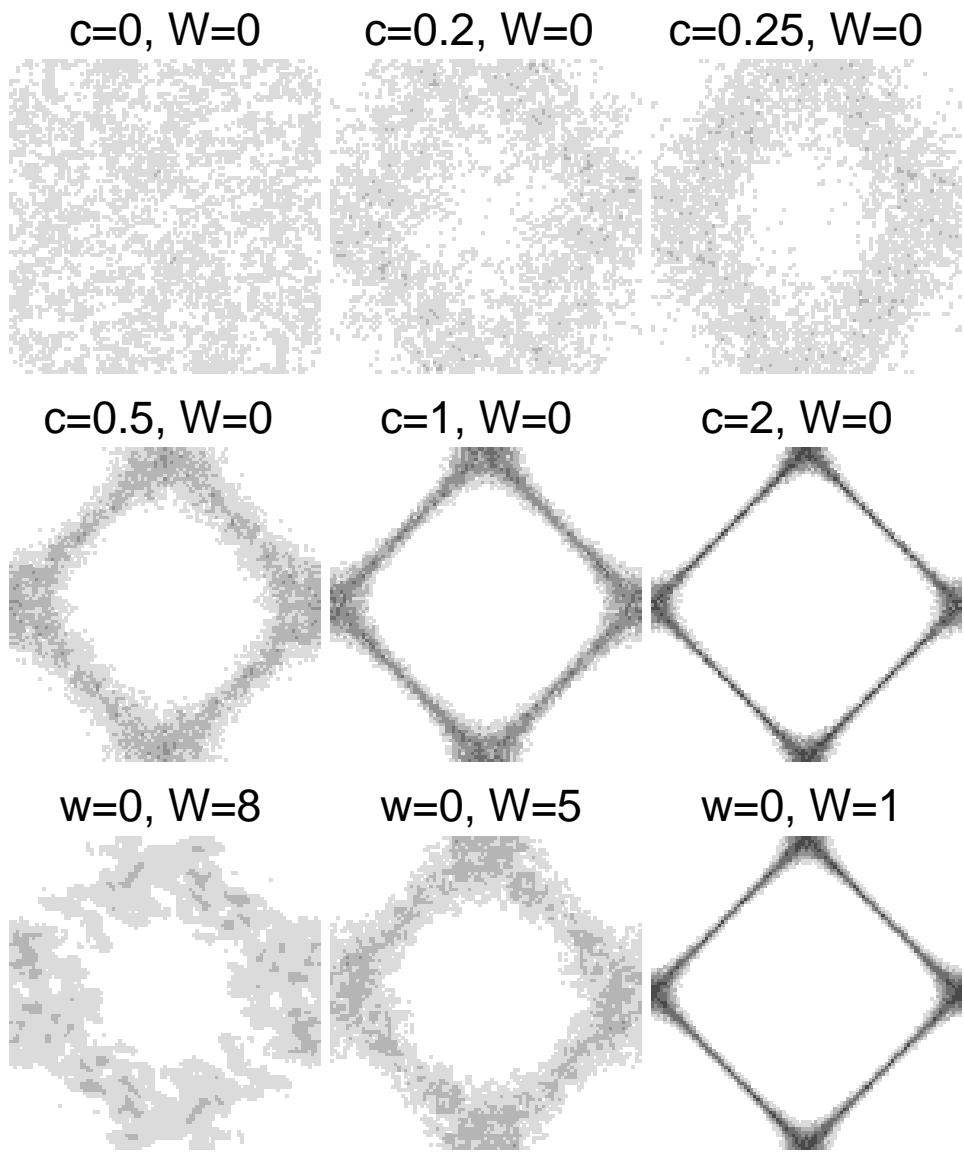,width=5.5in,height=6.4in}}
  \caption{
    Probability density of the Fourier transforms of eigenfunctions
    with the same parameters as shown in Fig.\ \protect\ref{fig-xphi},
    but averaged over $10$ states close to the band center. Different
    gray levels ($i=0,1,\ldots,6$) distinguish whether
    $|\phi_{j}(k_n,k_m)|^2 > 2^i/L^2$.}
\label{fig-kphi}
\end{figure}

According to the usual connection between real and Fourier space,
extended states in real space appear localized in Fourier space,
whereas localized states in real space appear extended in Fourier
space. Furthermore, eigenstates of the disordered system at energy $E$
are superpositions of eigenstates of the ordered system at energies
$E'= E \pm \Delta E(w,W)$, where $\Delta E(w,W)$ represents an energy
level broadening due to the disorder. For weak disorder the expansion
coefficients of this superposition are approximately equal for states
with small $\Delta E$. Interpreting $E$ as the Fermi energy, an
eigenstate of the weakly disordered system in Fourier space should
therefore exhibit the Fermi surface (FS).  Consequently, we can study
what happens to the FS upon increasing the disorder.
The 2D Fourier transform of the state $\phi_j(n,m)$ is defined as
\begin{equation}
 \label{FT}
  \phi_j(k_n,k_m) = \sum_{n}^{L} \sum_{m}^{L} \phi_j(n,m)
   \exp( \frac{2 \pi i k_n n}{L} )
   \exp( \frac{2 \pi i k_m m}{L} ).
\end{equation}
In Fig.\ \ref{fig-kphi}, we show probability densities of Fourier
transformed wavefunctions $\phi_j(k_n,k_m)$. As expected, weak
diagonal {\em and}\/ off-diagonal disorder produces states that appear
localized in Fourier space and reproduce the FS of the 2D
tight-binding model with nearest-neighbor hopping on a square lattice.
As examples consider the probability density plot at $c=2$ ($W=0$) and
the plot for $W=1$ ($w=0$).
With decreasing $c$ the states smear out, but the FS can still be
seen. Again, the behavior is qualitatively similar for both purely
off-diagonal and diagonal disorder as can be seen by comparing the
probability densities for $c=0.5$ ($W=0$) and $W=5$ ($w=0$) in Fig.\ 
\ref{fig-kphi}. The difference between the two types of disorder
appears only for $c < 0.5$. E.g., states for off-diagonal disorder $c
= 0$ appear completely delocalized in Fourier space, whereas for
strong diagonal disorder $W=8$ there is still a remnant of the FS.
This feature persists even at higher energies, suggesting different
localization properties of states in systems with off-diagonal
disorder characterized by small $c$.
%

%
%

\section{Density of States}
\label{sec-dos}

\begin{figure}
  \centerline{\psfig{figure=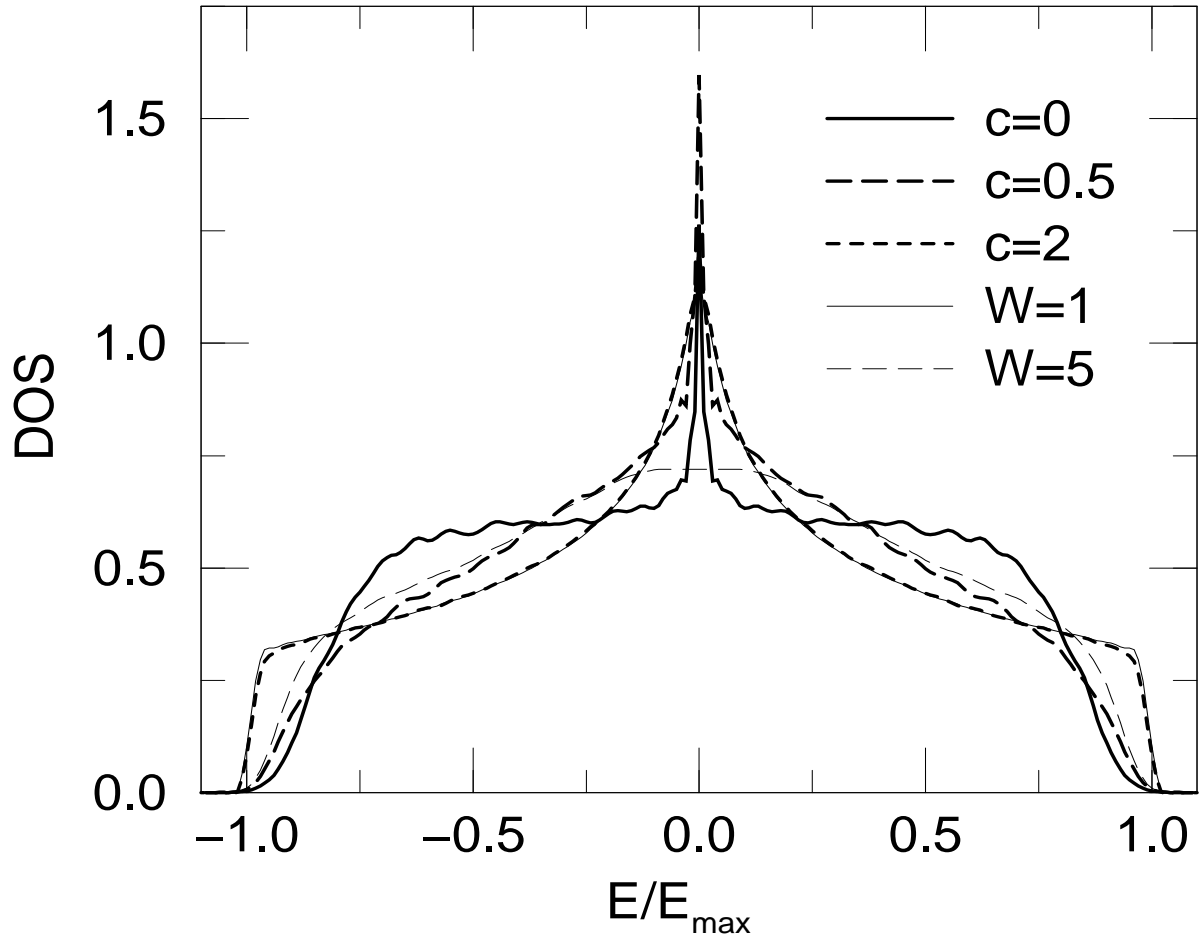,width=7.0in,height=5.5in}}
  \caption{
    Scaled density of states for purely off-diagonal disorder (thick
    lines, $W=0$) and, for comparison, purely diagonal disorder (thin
    lines, $w=0$). }
\label{fig-dos}
\end{figure}

In Fig.\ \ref{fig-dos}, we show the scaled DOS for off-diagonal
disorder obtained by averaging over many samples of size $N=96\times
96$. The off-diagonal disorder strengths are $c=0$, $0.5$, and $2$ with
$E_{max}=1.27$, $2.63$ and $8.24$, respectively.
For a 2D ordered system, the DOS has a logarithmic singularity at the
band center $E=0$. In the usual Anderson model with diagonal disorder,
this singularity is quickly suppressed when increasing the disorder
strength $W$ as shown in Fig.\ \ref{fig-dos} for $W=1$
($E_{max}=4.08$) and $W=5$ ($E_{max}=5.27$). Also, comparing in Fig.\ 
\ref{fig-dos} the DOS for weak off-diagonal disorder $c = 2$ ($W=0$)
and diagonal disorder $W=1$ ($w=0$), we see that both curves are
nearly identical.  However, diagonal and off-diagonal disorder are
qualitatively different for stronger disorders: Although the behavior
at the band edges is still similar, the peak at $E=0$ is more
pronounced for off-diagonal disorder $c = 0.5$, while the
diagonal-disorder case $W=5$ does not show any such singularity.
It therefore appears that it is in the band center $E=0$ where any
differences between purely diagonal as compared to purely off-diagonal
disorder are likely to be most relevant.

%
%

\section{Localization properties of the eigenstates}
\label{sec-locprop}

Thus far we have only qualitatively studied the difference of diagonal
and off-diagonal disorder with respect to the localization properties.
In the present chapter, we will investigate the localization
properties quantitatively by an analysis of the participation numbers
and the multifractal characteristics.


\subsection{Participation numbers}
\label{subsec-partnum}

Let $\phi_j(n,m)$ denote the wave function amplitude of the $j$th
normalized eigenstate at site $(n,m)$. A simple measure of the number
of sites which contribute to this wave function is the participation
number $P_N(j)$. It is defined as
\begin{equation} \label{partno}
  P_{N}^{-1}(j) = \sum_{n,m} |\phi_{j}(n,m)|^4.
\end{equation}
Thus a completely localized state $\phi_j(n,m)= \delta_{n_0,n}
\delta_{m_0,m}$ corresponds to $P_N = 1$, whereas a fully extended
state $\phi_j(n,m)= 1/\sqrt{N}$ has $P_N = N$.

\begin{figure}
  \centerline{\psfig{figure=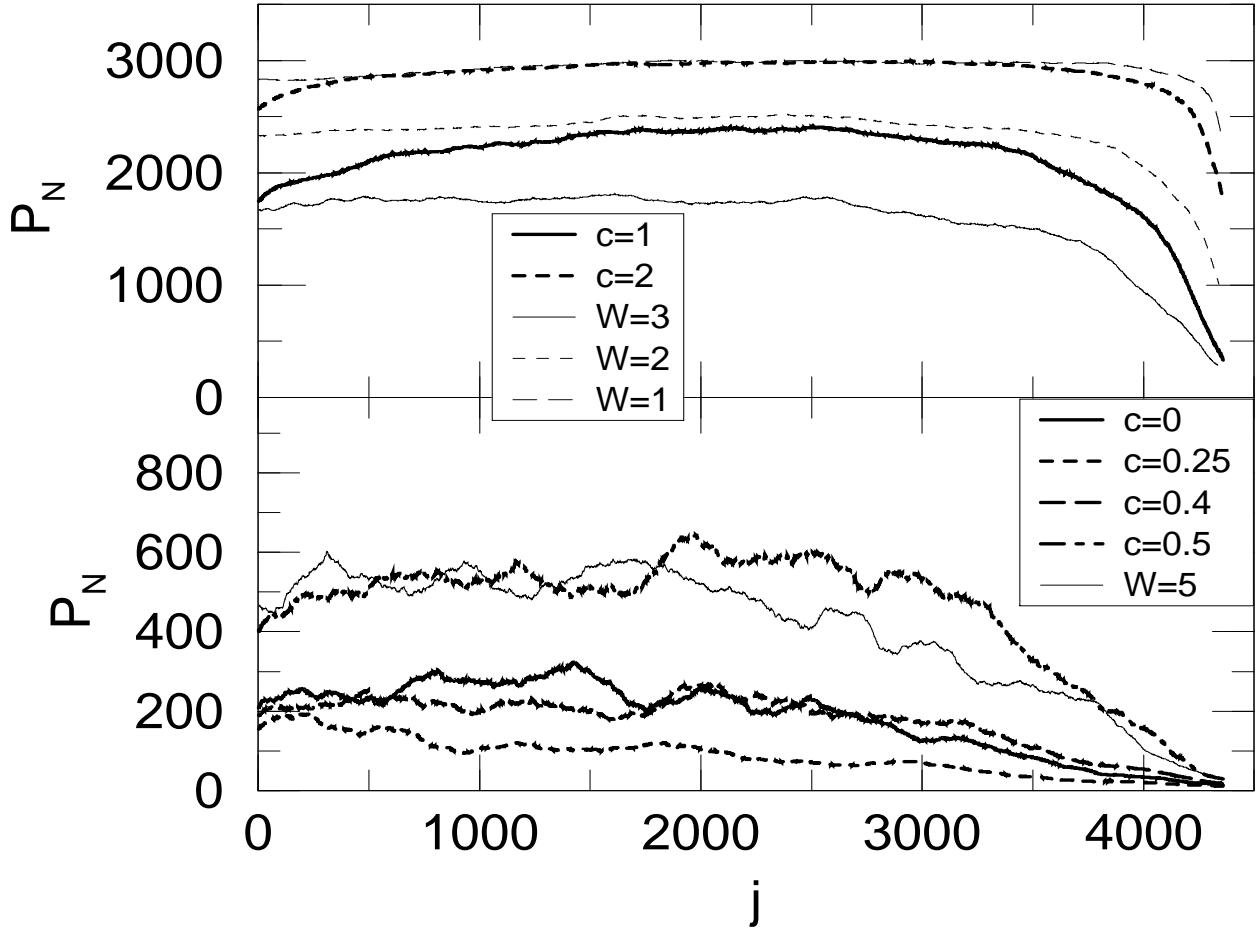,width=7.0in,height=5.5in}}
  \caption{
    Averaged participation numbers $P_N$ versus the number $j$ of the
    eigenstate ordered with increasing energy ($0 \leq E_j \leq
    E_{j+1}$) and $N=96\times 96$. Purely off-diagonal disorders are
    shown by thick lines, purely diagonal disorders by thin lines.}
\label{fig-partnum}
\end{figure}

Figure \ref{fig-partnum} shows the changes of the participation
numbers within the band.  As the $P_N$ values for neighboring states
exhibit large fluctuations a moving average over 250 consecutive
states was applied to the data to produce smoother curves. We first
note that as observed in sections \ref{sec-wavefcn} and \ref{sec-dos},
the behavior for weak disorder $c=2$ ($W=0$) and $W=1$ ($w=0$) and
also for stronger disorder $c=0.5$ ($W=0$) and $W=5$ ($w=0$) is again
similar.
For all disorders, both diagonal as well as off-diagonal, $P_N$
decreases at the band edges, where one expects the strongest
localization of states. Differences between diagonal and off-diagonal
disorder occur close to the band center. For weak off-diagonal
disorder a minimum of $P_N$ at $E=0$ is well pronounced, whereas no
such feature exists in diagonally disordered systems. For stronger
disorder the $P_N$ values show large fluctuations. Still, we observe
that the $P_N$ values decrease close to the band center for all values
of $c$. Thus we are led to the preliminary hypothesis that the peak in
the DOS at $E=0$ for off-diagonal disorder corresponds to states which
are more strongly localized than states at small but finite energies
away from the band center.

A further interesting conclusion regarding the off-diagonal disorder
strength $c$ may be drawn from Fig.\ \ref{fig-partnum}. While
decreasing $c$ results, as expected, in stronger localization, we
nevertheless observe the strongest disorder effect {\em not}\/ for $c
= 0$. Rather, the value of $c$ at which we observe the smallest $P_N$,
and thus the strongest localization, can be located around $c = 0.25$
just as in section \ref{sec-wavefcn}. The $P_N$ values corresponding
to $c = 0$ are larger and approximately the same as for $c = 0.4$.


\subsection{Multifractal analysis}
\label{subsec-mfa}

Another useful tool for the characterization of the eigenstates of
disordered systems in 2D is the multifractal analysis \cite{mfa}.  As
is immediately clear from Fig.\ \ref{fig-xphi}, the simple notions of
exponentially localized or homogeneously extended states are
invalidated by large fluctuations of the probability density --- at
least at small length scales. It has been shown in recent studies
\cite{mfarec} of the Anderson Hamiltonian with diagonal disorder that
its eigenstates have multifractal characteristics which are related to
their localization properties.
Our multifractal analysis of the eigenfunctions is based on the
standard box-counting procedure \cite{boxc}: We divide the $N= L
\times L$ lattice into a number of ``boxes'' of size $\delta
L\times\delta L$. We then determine the contents
$\mu_{i}(\delta)=\sum_{(n,m)\in i} | \phi(n,m) |^2$ of each box $i$
for a given eigenfunction $\phi(n,m)$. The normalized $q$th moment of
the box probability $\mu_i(q,\delta)=
\mu_{i}^{q}(\delta)/\sum_{k}\mu_{k}^{q}(\delta)$ constitutes a measure
and may be used to define the singularity strength (Lipschitz-H\"older
exponent)
\begin{equation} \label{eq-LHexp}
 \alpha(q) = \lim_{\delta \rightarrow 0}
             \sum_{i} \mu_i(q,\delta) \ln \mu_i(1,\delta) / \ln\delta
\end{equation}
and the corresponding fractal dimension
\begin{equation} \label{eq-fracdim}
f(q) = \lim_{\delta \to 0} \sum_{i} \mu_i(q,\delta)
       \ln \mu_i(q,\delta) / \ln \delta.
\end{equation}
We plot the sums in Eqs.\ (\ref{eq-LHexp}) and (\ref{eq-fracdim})
versus $\ln \delta$ and observe multifractal behavior if and only if
the data may be fitted well by straight lines for small $\delta$. This
is indeed the case for our data and the slopes from the linear
regression procedure used in the fits give the singularity spectrum
$f(\alpha)$. We emphasize that a check on the linearity is important,
since the numerical procedure gives an $f(\alpha)$ curve for nearly
every distribution of the local probability densities, but without the
linearity it does not indicate multifractality.
From Eqs.\ (\ref{eq-LHexp}) and (\ref{eq-fracdim}) one can obtain a
set of generalized dimensions $D(q)= \lbrace f[ \alpha(q)] - q
\alpha(q) \rbrace / (1-q)$. Then $D(0)$ is simply the Hausdorff
dimension of the underlying support (and thus $2$ in the 2D case),
$D(1)=\alpha(1)=f(1)$ gives the entropy or information dimension and
$D(2)$ represents the correlation dimension \cite{mfa}.

\begin{figure}
  \centerline{\psfig{figure=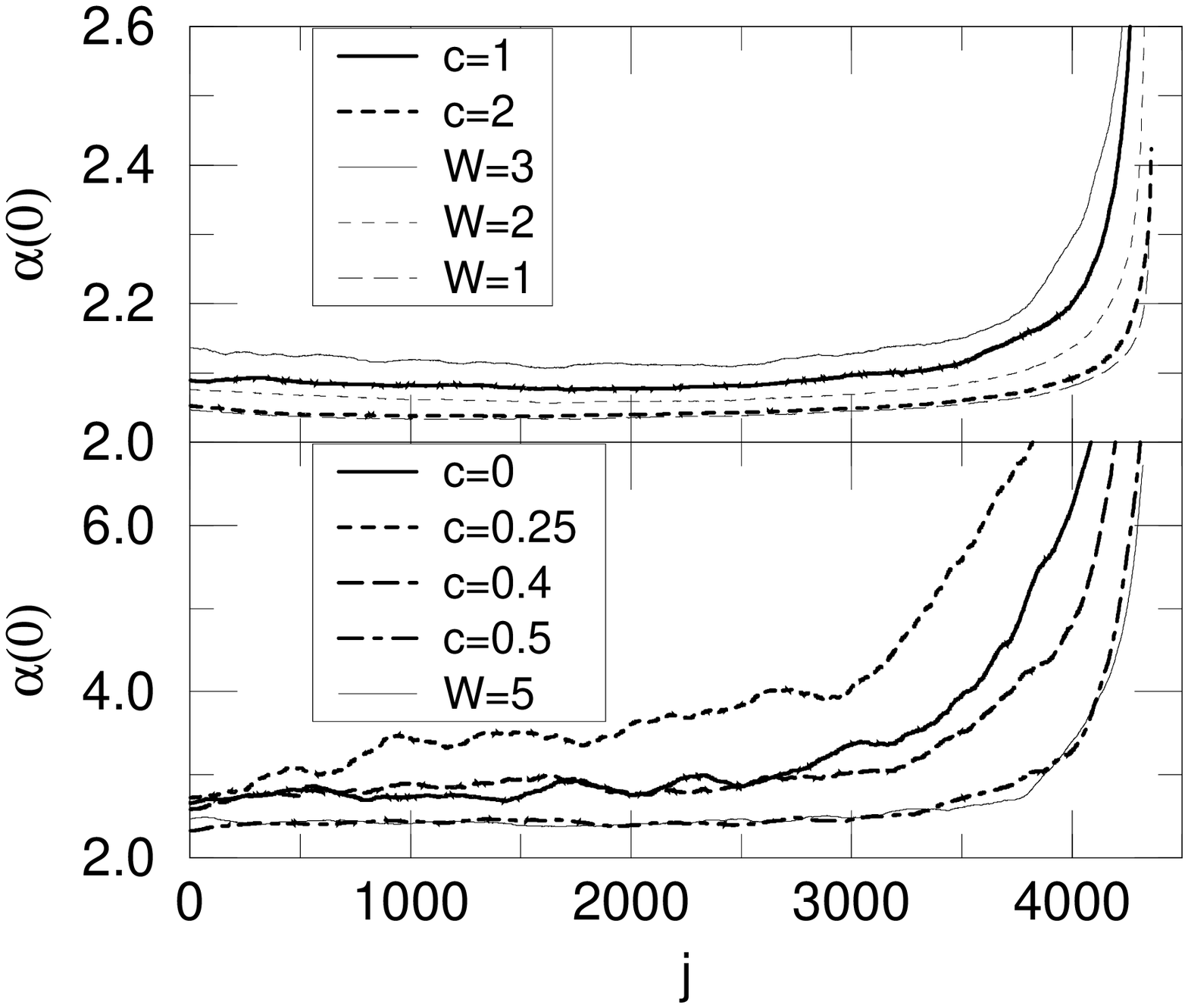,width=7.0in,height=5.5in}}
  \caption{
    Singularity strength $\alpha(0)$ versus the number $j$ of the
    eigenstate ordered with increasing energy ($0 \leq E_j \leq
    E_{j+1}$) and $N=96\times 96$.  Purely off-diagonal disorders are
    shown by thick lines, purely diagonal disorders by thin lines.}
\label{fig-a0}
\end{figure}

\begin{figure}
  \centerline{\psfig{figure=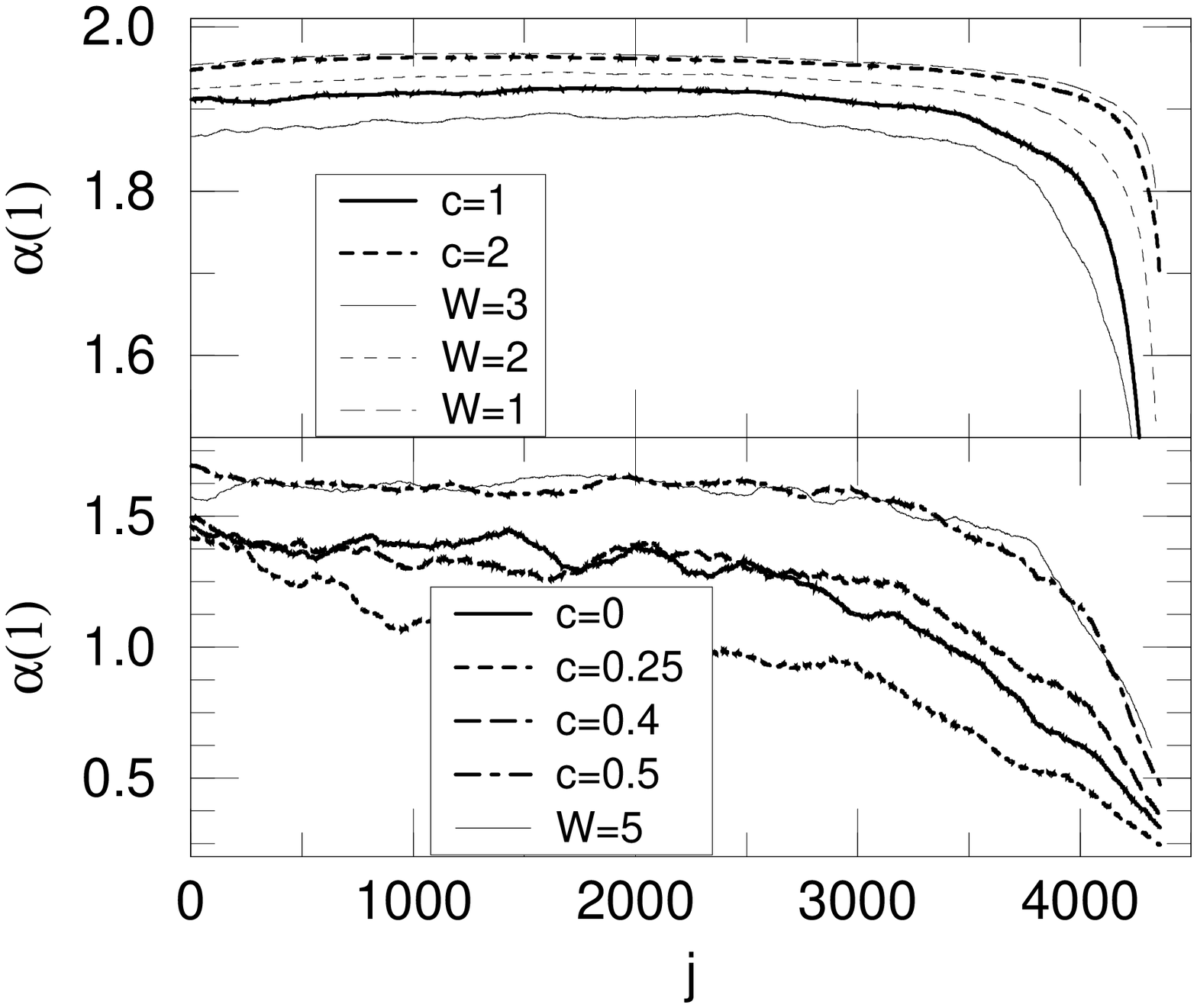,width=7.0in,height=5.5in}}
  \caption{
    Singularity strength $\alpha(1)$ versus the number $j$ of the
    eigenstate ordered with increasing energy ($0 \leq E_j \leq
    E_{j+1}$) and $N=96\times 96$.  Purely off-diagonal disorders are
    shown by thick lines, purely diagonal disorders by thin lines.}
\label{fig-a1}
\end{figure}

For a truly extended 2D wave function, $\alpha(q)=f(q)=2$.  The more a
state becomes localized, the more the values differ from $2$. We show
in Figs.\ \ref{fig-a0} and \ref{fig-a1} the calculated values for
$\alpha(0)$ and $\alpha(1)$, respectively. Again, moving averages over
250 states are determined. The deviations from $2$ are well pronounced
at the band edge, where $\alpha(0)$ increases and $\alpha(1)$
decreases drastically.  Therefore localization of states at the band
edge is confirmed by fractal measures in agreement with the above
results from participation numbers.
If, as suggested by participation numbers in the last section for the
off-diagonal disorder, localization would increase at the band center,
we should expect a similar deviation of the $\alpha$ values from $2$,
while there should be no significant change for diagonal disorder.
However, the differences between the $\alpha$ values are negligible
for both weak disorders $W$ and $c$. For stronger off-diagonal
disorder even the opposite tendency can be observed: the $\alpha$
values tend towards $2$, which suggests rather a tendency towards
weaker localization. Similar results can be found from, e.g., $D(2)$
and $f(\alpha)$. Without showing the plots, we only note that the
values of $D(2)$ for purely off-diagonal disorder in the band center
are close to $1$ for $c \leq 0.5$.

Despite of this apparent disagreement with section
\ref{subsec-partnum} --- which we will resolve in the next subsection
--- the fractal characteristics clearly confirm the previous
observations that the strongest off-diagonal disorder appears for $c =
0.25$ and the $\alpha$ values for the disorders $c = 0$ and $c = 0.4$
are close, indicating a similarity of the localization properties.


\subsection{Scaling of the participation numbers}
\label{subsec-scapartnum}

The above mentioned disagreement between the localization properties
at the band center derived from participation numbers and multifractal
characteristics may be understood by taking into account that for a
given system size $N$, the $P_N$ values do not reflect directly the
localization of the state in the infinite system. One should rather
look at the dependence of $P_N$ on $N$, since $P_N$ scales with $N$ as
\begin{equation}
  P_N \sim N^{\kappa}.
\end{equation}
Thus for a localized state $\kappa=0$, whereas for an extended state
$\kappa=1$. The connection to the multifractal properties of the last
section is given by the relation $P_{N} \sim N^{D(2)/D(0)}$
\cite{wegener2}.

\begin{figure}
  \centerline{\psfig{figure=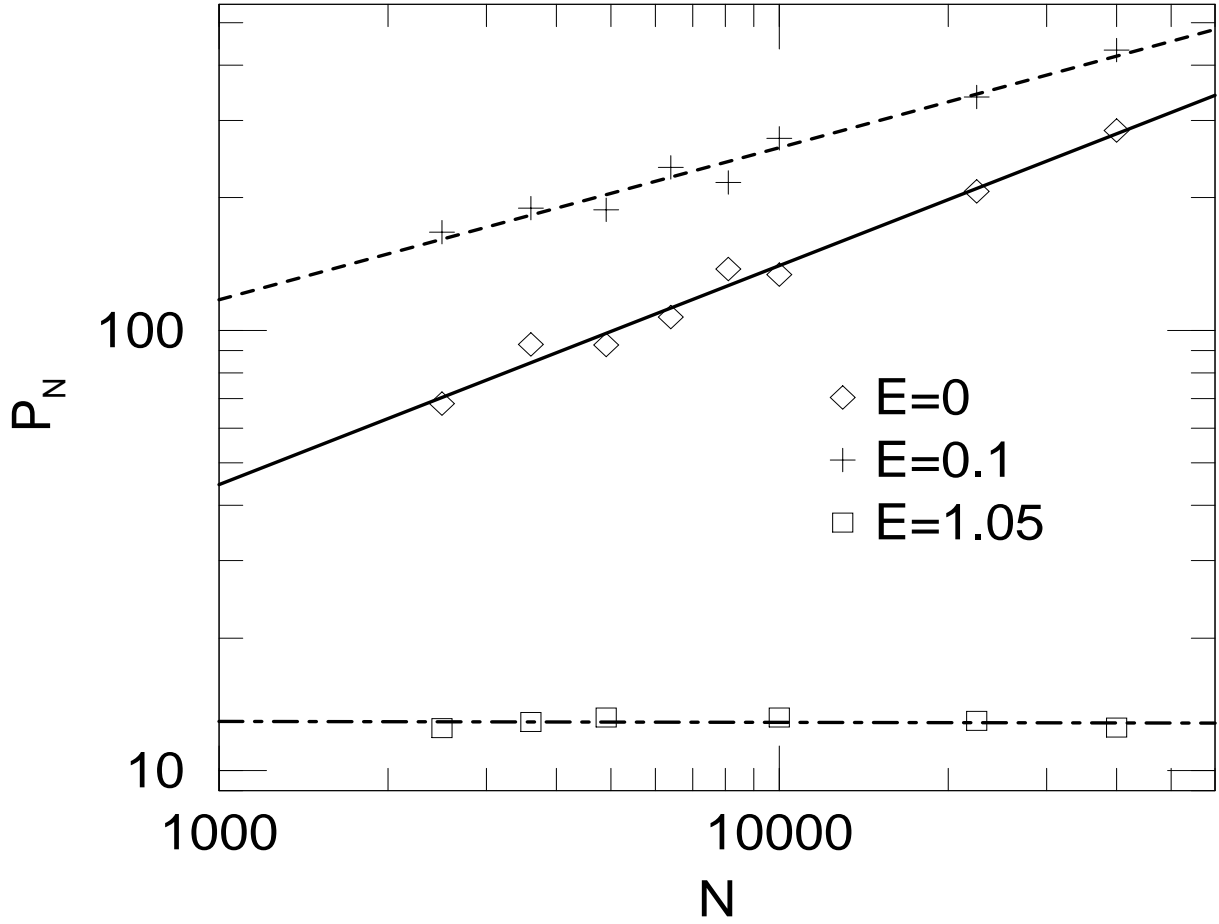,width=7.0in,height=5.5in}}
  \caption{
    Finite-size dependence of the participation numbers $P_N$ for
    eigenstates with $c=0$ at energies in the band center ($E=0$),
    outside the band center but still close to the peak in the DOS
    ($E=0.1$), and close to the band edge ($E=1.05$).}
\label{fig-scapartnum}
\end{figure}

In Fig.\ \ref{fig-scapartnum}, we show the dependence of $P_N$ on $N$
for off-diagonal disorder with $c = 0$ for system sizes up to $N= 200
\times 200$.  The $P_N$ data were averaged over different disorder
realizations and over a small energy interval $\Delta E = 0.0005$ for
$E=0$ and $E=0.1$ or $\Delta E=0.01$ for $E=1.05$.  The latter
interval is larger due to the small DOS close to the band edge.  The
number of states taken into averaging was about $100$.  A
least-squares fit gives the slope of the straight line in the log-log
plot; $\kappa=0.00\pm 0.03$ close to the band edge at $E=1.05$,
$\kappa=0.34\pm 0.06$ for $E=0.1$ and $\kappa=0.50\pm 0.06$ for $E=0$
in agreement with the value of $D(2)$ obtained in the last section.

The result $\kappa=0$ suggests again that the states at the band edge
are completely localized and the $P_N$ constant.  The numerical values
of $P_N$ are also the smallest in this energy range. Although the
values of $P_N$ for $E=0$ are smaller than for $E=0.1$, suggesting
stronger localization as in section \ref{subsec-partnum}, $\kappa$ is
{\em bigger} at the band center which means that the state is {\em
  less} localized.
In fact, $\kappa=0.5$ is far away from the localized behavior
$\kappa=0$, but also from the $\kappa=1$ value of extended states.
This suggests that while the state at $E=0$ is clearly not extended,
it may have properties similar to {\em critical} states, i.e.\ states
at the MIT. This is corroborated by the observation \cite{mfarec} that
at the MIT in the 3D isotropic and anisotropic Anderson models one
finds $D(2)$ values in the range $[1.2,1.6]$. Also, for the Anderson
model defined on two bifractals \cite{grussphd} one finds $D(2)\approx
1.98$ and $2.07$ with $D(0)= 2.58$.  Thus $\kappa=D(2)/D(0)$ for
critical states is typically in the range $[0.4, 0.8]$ and we propose
that the value $\kappa=0.5$ in the present case indicates a
delocalization-localization transition.  We emphasize, however, that
the non-zero slope for $E=0$ may be a finite-size effect and the $P_N$
curves may bend down for $N> 200 \times 200$ and eventually even
become flat.


\subsection{The strongest off-diagonal disorder}
\label{subsec-strongod}

As we have shown above, the strongest tendency towards localization
appears for $c = 0.25$ and not, as one might expect, for $c = 0$. This
may be rationalized as follows: the strength of the disorder is the
larger the broader the distribution $P(t)$ of the off-diagonal hopping
elements is when compared to the mean value of the hopping element,
i.e., the larger the ratio $w/c$ is.  There is, however, yet another
factor which should be taken into account. The localization of the
eigenstates should be more pronounced when more hopping elements are
close to $0$, because a small hopping stops the propagation of the
electrons across the system. This effect is related to the
distribution $P(|t|)$ of the absolute values of the hopping elements.
Its importance can be described by the ratio of the mean value of
$P(|t|)$ to the variance, which reaches its minimum close to $c =
0.4$. Thus we may expect the largest obstruction of the
propagation of an electron wave function at $c\approx 0.4$. 
The overall effect of the hopping disorder is a combination of the
width of $P(t)$ and $P(|t|)$. As shown in the last sections, it is
most pronounced between $c = 0$ and $c = 0.4$.  In fact the maximum
effect seems to be reached at about $c = 0.25$.  This is also
consistent with the observed similarity between system with disorder
$c = 0$ and $c \approx 0.4$.

%
%

\section{Calculation of localization lengths}
\label{sec-fssloclen}

In the previous section, we have shown that the state at $E=0$ for the
Anderson Hamiltonian with purely off-diagonal disorder may by
characterized both by the system size dependence of the participation
numbers and by its multifractal properties as being similar to
critical states observed at the MIT in the higher-dimensional Anderson
models with diagonal disorder \cite{mfarec}. In this section, we will
confirm this characterization by an independent numerical method and
also study the stability of the state with respect to an {\em
  additional} potential disorder $W$.


\subsection{The transfer-matrix method}
\label{subsec-tmm}

Perhaps the most suitable method to directly assess localization
properties of states for non-interacting disordered systems is the
calculation of the decay lengths of wave functions on quasi-1D strips
of width $M$ and length $K \gg M$ by means of the TMM
\cite{pichard,mckinnon}. To this end, the Schr\"{o}dinger equation is
written as
\begin{equation}
 \label{eq-schr}
 t^{||}_{n+1,m} \psi_{n+1,m}
 =(E-\epsilon_{n,m})\psi_{n,m} -
  t^{\perp}_{n,m+1}\psi_{n,m+1} -
  t^{\perp}_{n,m}\psi _{n,m-1} -
  t^{||}_{n,m}\psi_{n-1,m},
\end{equation}
where $\psi_{n,m}$ is the wave function at site $(n,m)$,
$t^{\perp}_{n,m}$ represents the hopping element from site $(n,m)$ to
site $(n,m - 1)$ and $t^{||}_{n,m}$ represents the hopping element
from $(n-1,m)$ to $(n,m)$. Equation (\ref{eq-schr}) may be
reformulated in the TMM form as
\begin{equation}
\label{eq-tmm}
\left(
\begin{array}{c}
\psi _{n+1} \\
\psi _n
\end{array}
\right) =\left(
\begin{array}{cc}
[t^{||}_{n+1}]^{-1}(E-\epsilon_n - H_{\perp}) & - [t^{||}_{n+1}]^{-1}t^{||}_{n} \\
1 & 0
\end{array}
\right) \left(
\begin{array}{c}
\psi _n \\
\psi _{n-1}
\end{array}
\right) =T_n\left(
\begin{array}{c}
\psi _n \\
\psi _{n-1}
\end{array}
\right),
\end{equation}
where $\psi_n= (\psi_{n,1}, \psi_{n,2}, \ldots, \psi_{n,M})^T$ denotes
the wave function at all sites of the $n$th slice, $\epsilon_n= {\rm
  diag}(\epsilon_{n,1}, \ldots, \epsilon_{n,M})$, $H_{\perp}$ the
hopping Hamiltonian within slice $n$ and $t^{||}_n=$ ${\rm
  diag}(t^{||}_{n,1}$, $t^{||}_{n,2}$, $\ldots$, $t^{||}_{n,M})$ the
diagonal matrix of hopping elements connecting slice $n-1$ with slice
$n$. The evolution of the wave function is given by the product of the
transfer matrices $\tau_K = T_K T_{K-1} \ldots T_2 T_1$. According to
Oseledec's theorem \cite{Oseledec} the eigenvalues $\exp [\pm\gamma
_i(M)]$ of $\Gamma=\lim_{K\rightarrow \infty
  }(\tau_K^{\dagger}\tau_K)^{1/2K}$ exist and the smallest Lyapunov
exponent $\gamma_{min}>0$ determines the largest localization length
$\lambda(M)=1/ \gamma_{min}$ at energy $E$. The accuracy of the
$\lambda$'s is determined as outlined in
Ref.\ \cite{mckinnon} from the variance of the changes of the
exponents in the course of the iteration.

For $c \leq 0.5$, there is always a small probability that one of the
$t^{||}_{n,m}$ is close to $0$ such that a division as prescribed
above may lead to numerically unreliable results. We have therefore
applied a cutoff for small $|t^{||}_{n,m}|$ and checked that our
$\gamma_{min}$ values are independent of the cutoff.

According to the one-parameter-scaling hypothesis \cite{go4,mckinnon},
the reduced localization lengths $\lambda(M)/M$ for different
disorders and energies scale onto a single scaling curve, i.e.,
\begin{equation}
  \lambda(M)/M = f(\xi /M).
\end{equation}
As usual, we determine the finite-size-scaling (FSS) function $f$ and
the values of the scaling parameter $\xi$ by a least-squares fit and
the absolute scale of $\xi$ can be obtained by fitting $\lambda/M =
\xi/M + b (\xi/M)^2$ for the smallest localization lengths
\cite{mckinnon}.  For diagonal disorder in 2D, this hypothesis has
been shown to be valid with very high accuracy, and only one branch of
the scaling curve $f$ exists which corresponds to localized behavior
\cite{pichard,mckinnon}. Furthermore, the $\xi$ values of this branch
are just equal to the localization length in the infinite system.


\subsection{Off-diagonal disorder}
\label{subsec-pod}

The TMM calculations for purely off-diagonal disorder ($W=0$) have
been performed with at least $1\%$ accuracy for different $c$ values.
In order to achieve this accuracy, we needed substantially more
transfer-matrix multiplications as for diagonal disorder.

\begin{figure}
  \centerline{\psfig{figure=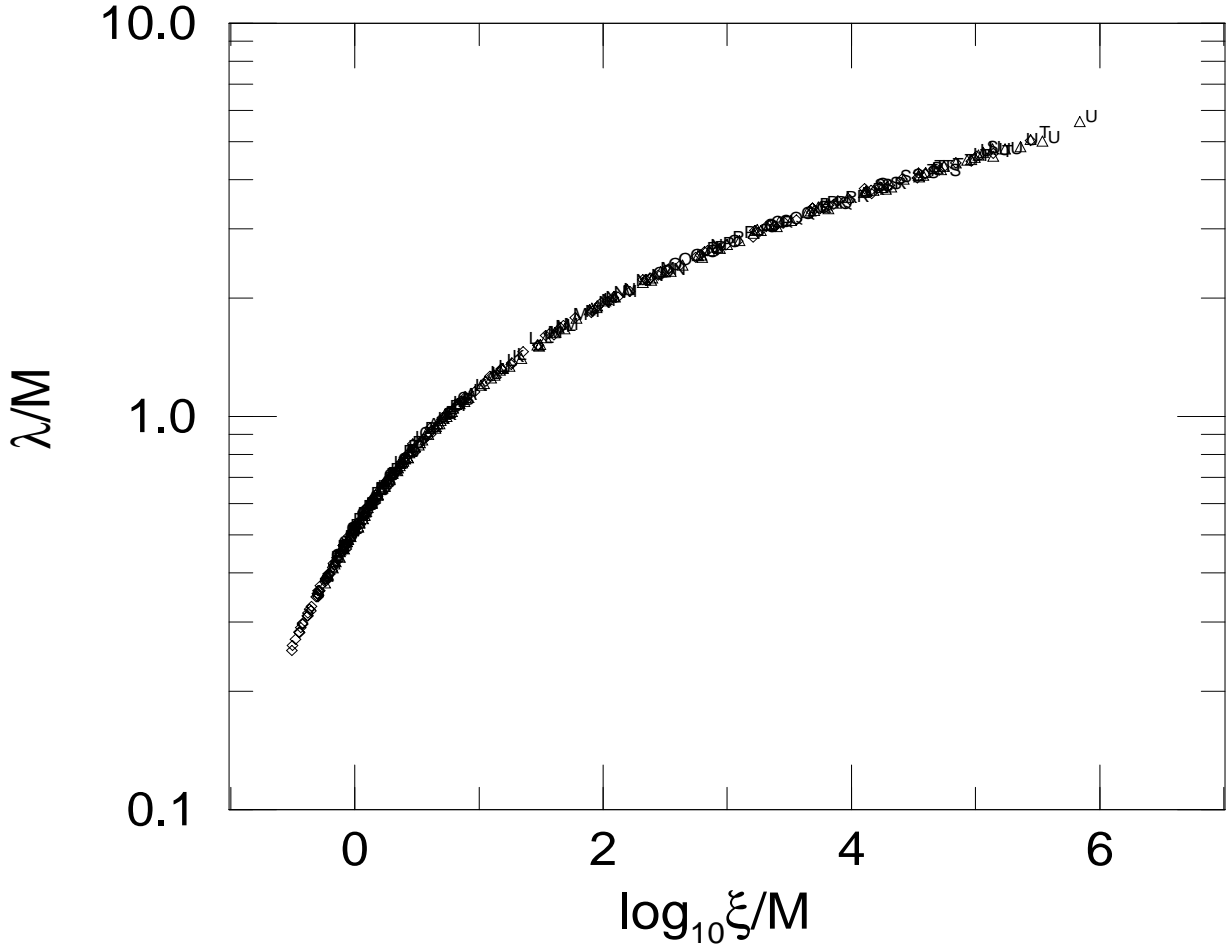,width=7.0in,height=5.5in}}
  \caption{
    Finite-size scaling plot of the reduced localization lengths
    $\lambda(M)/M$ for purely random hopping ($W=0$) outside the band
    center with energies $E= 0.005$ (characters), $E=0.01$
    ($\triangle$) and $E=0.1$ ($\Diamond$). The off-diagonal disorder
    strengths corresponding to $c= 0, 0.05, \ldots, 1$ are indicated
    by A, B, $\ldots$, U, respectively.}
\label{fig-fss-offdiag1}
\end{figure}

\begin{figure}
  \centerline{\psfig{figure=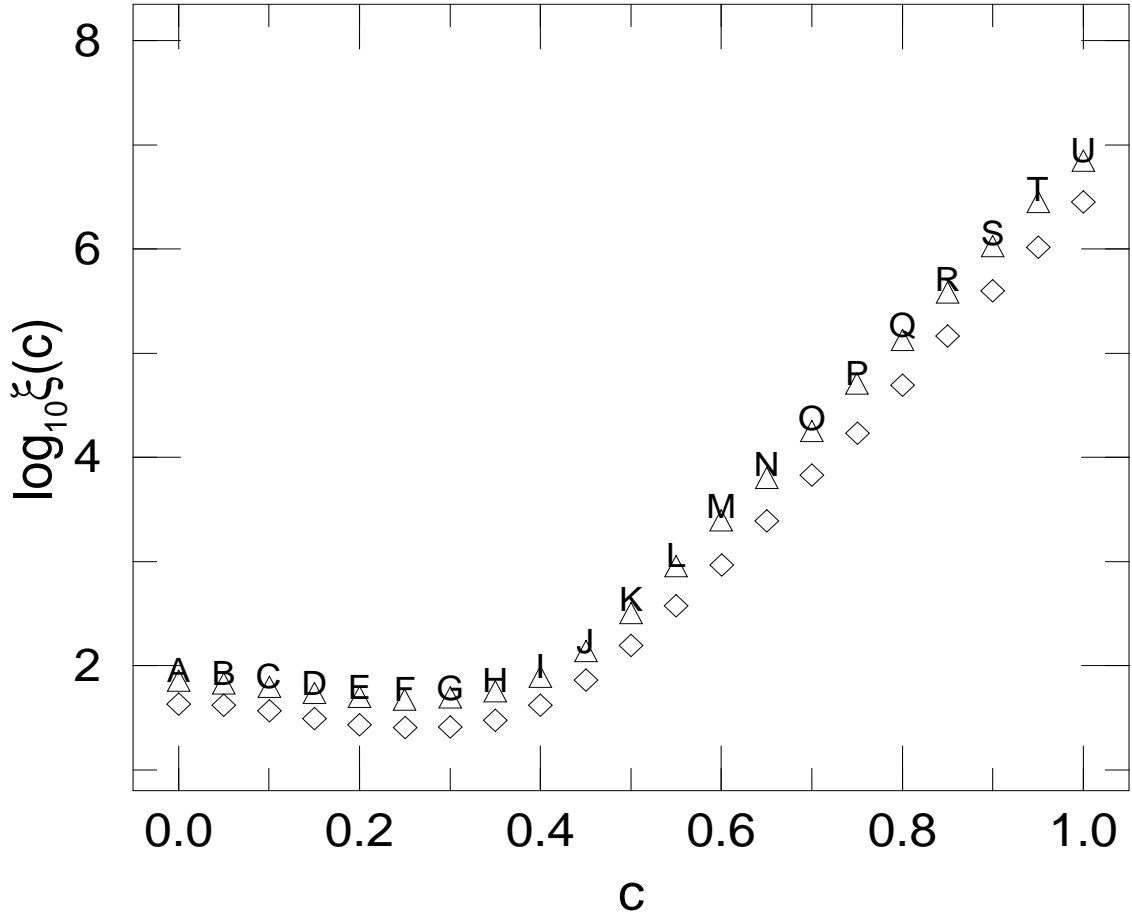,width=7.0in,height=5.5in}}
  \caption{
    Scaling parameter $\xi$ as a function of off-diagonal disorder
    center $c$ for $W=0$ and $E= 0.005$ (characters), $0.01$
    ($\triangle$) and $0.1$ ($\Diamond$).}
\label{fig-sca-offdiag}
\end{figure}

The FSS results for the localization lengths obtained by the TMM for
off-diagonal disorder of $w=1$ with $c$ values ranging from $0$ to $1$
and energies {\em outside} the band center are displayed in Fig.\ 
\ref{fig-fss-offdiag1}. The strip widths were $M=10, 20, \ldots, 80$.
As can be seen, the reduced localization length $\lambda(M)/M$ can be
scaled onto a single curve for all $c$ and $E$, thus confirming the
validity of the scaling hypothesis also for purely off-diagonal
disorder.  Moreover, we obtain only one branch of the scaling function
corresponding to localization.
In Fig.\ \ref{fig-sca-offdiag}, we show the dependence of the scaling
parameter $\xi$ on $c$. It exhibits a minimum close to $c=0.25$. This
shows in agreement with section \ref{sec-locprop} that the maximum
strength of the off-diagonal disorder appears for $c=0.25$.  The
disorders with $c = 0.4$ and $c = 0$ have approximately the same
strength.

\begin{figure}
  \centerline{\psfig{figure=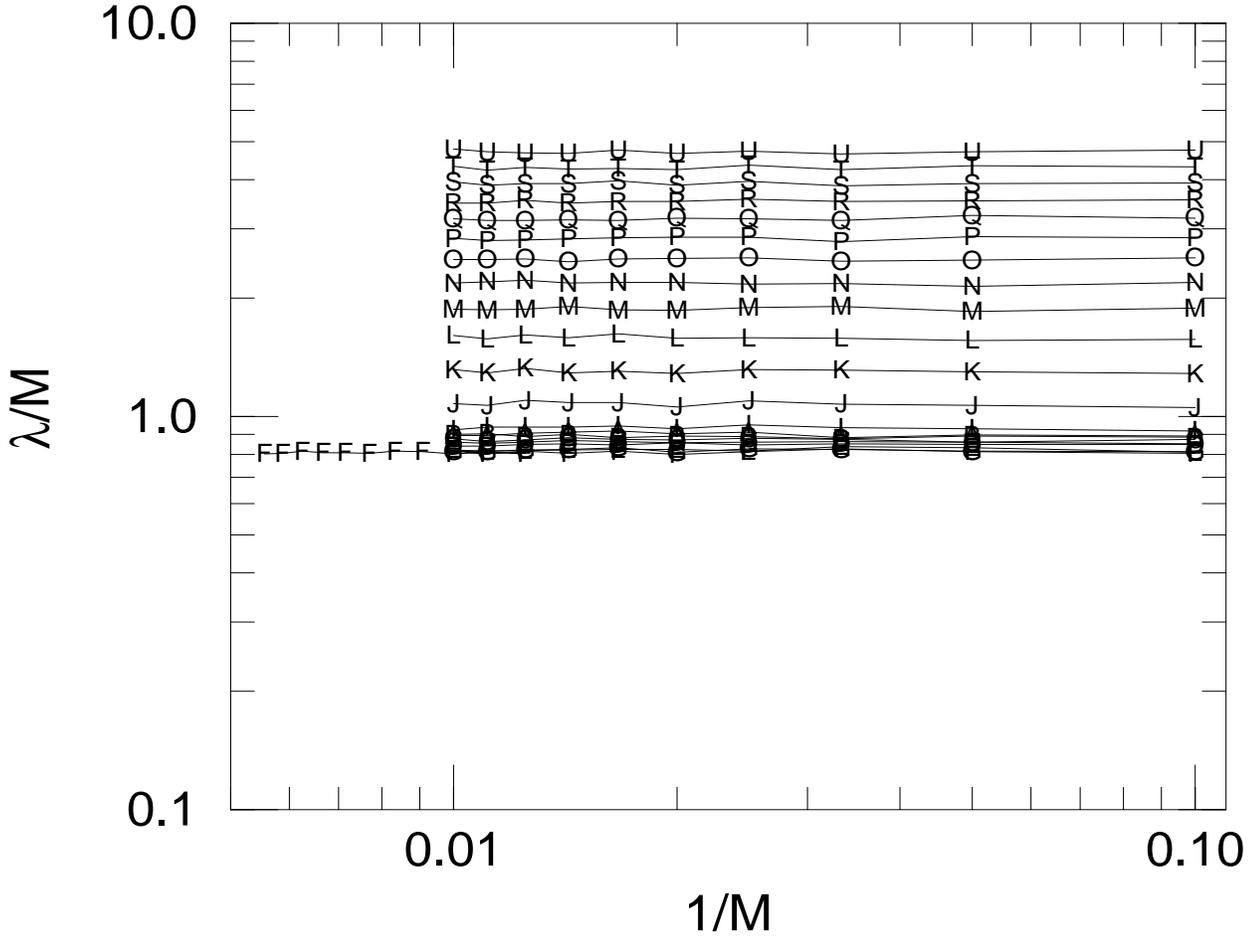,width=7.0in,height=5.5in}}
  \caption{
    Reduced localization lengths $\lambda(M)/M$ vs.\ $1/M$ for purely
    random hopping ($W=0$) and $E = 0$. The characters represent
    different $c$ values as in Figs.\ \protect\ref{fig-fss-offdiag1}
    and \protect\ref{fig-sca-offdiag}. The $\lambda/M$ scale is the
    same as in Fig.\ \protect\ref{fig-fss-offdiag1} for easier
    comparison.  All curves are parallel to the $1/M$-axis even up to
    $M=180$ for $c = 0.25$ (F).}
\label{fig-fss-offdiag2}
\end{figure}

We now turn to the state at $E=0$. As shown in Fig.\ 
\ref{fig-fss-offdiag2}, the reduced localization lengths $\lambda /M$
are constant vs.\ $1/M$. The curves for different $c$ do not overlap
and FSS is impossible. This is typical for the {\em critical}\/
behavior observed at the MIT in the 3D Anderson model \cite{mckinnon}.
For the strongest off-diagonal disorder $c = 0.25$, we have used strip
widths up to $M=180$. Still, there is no bending down in the curve
which suggests the persistence of criticality up to these rather large
$M$.  In addition to the periodic boundary conditions used so far, we
have also considered the TMM problem (\ref{eq-tmm}) with hard-wall and
aperiodic boundary conditions.  Although the actual values of the
localization lengths differ slightly, the behavior remains critical up
to $M=180$. In view of the particle-hole symmetry mentioned in section
\ref{sec-model}, we note that these results hold equally well for $M$
odd.
We emphasize that the presence of the critical state is restricted to
$E=0$ for all off-diagonal disorders.  All calculations for larger
energies indicate localized states only.  Note, e.g., that states for
$E=0.005$ and small $c$ belong already to the peak in the DOS of Fig.\ 
\ref{fig-dos}. Nevertheless, they are clearly localized as shown in
Figs.\ \ref{fig-fss-offdiag1} and \ref{fig-sca-offdiag}.


\subsection{Additional diagonal disorder}
\label{subsec-add}

\begin{figure}
  \centerline{\psfig{figure=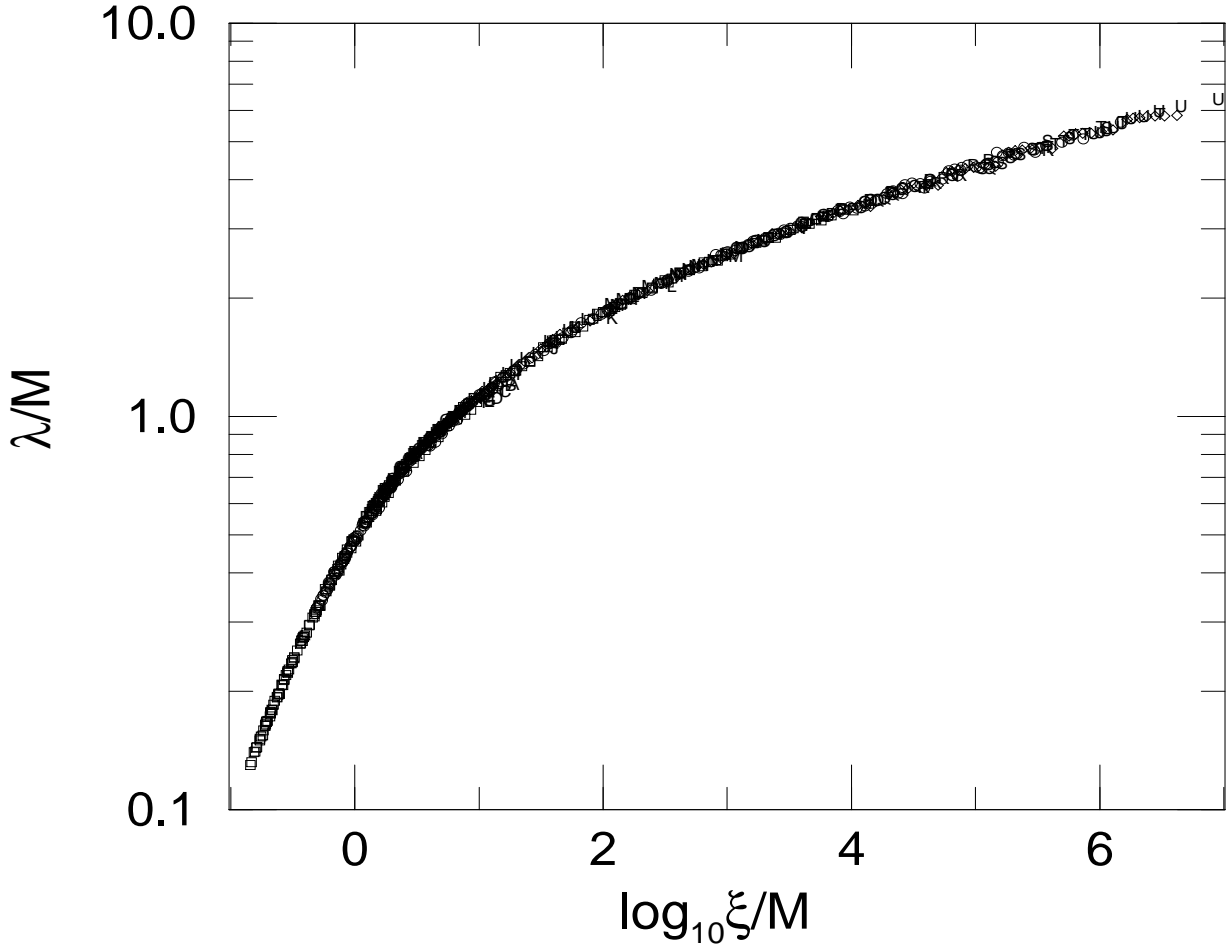,width=7.0in,height=5.5in}}
  \caption{
    Finite-size scaling plot of the reduced localization lengths
    $\lambda(M)/M$ for random hopping at $E=0$ and additional
    potential disorder $W=0.0001$ ($\diamond$), $0.001$ (characters as
    in Fig.\ \protect\ref{fig-fss-offdiag1}), $0.01$ ($\circ$) and
    $0.1$ ($\Box$). For $W=0.0001$, only data with $M\geq 40$ has been
    used. The small deviations from FSS at $\xi/M\approx 10$,
    $\lambda/M\approx 1$ are coming from data for $W=0.001$ and
    $M=10$.}
\label{fig-fss-diag1}
\end{figure}

\begin{figure}
  \centerline{\psfig{figure=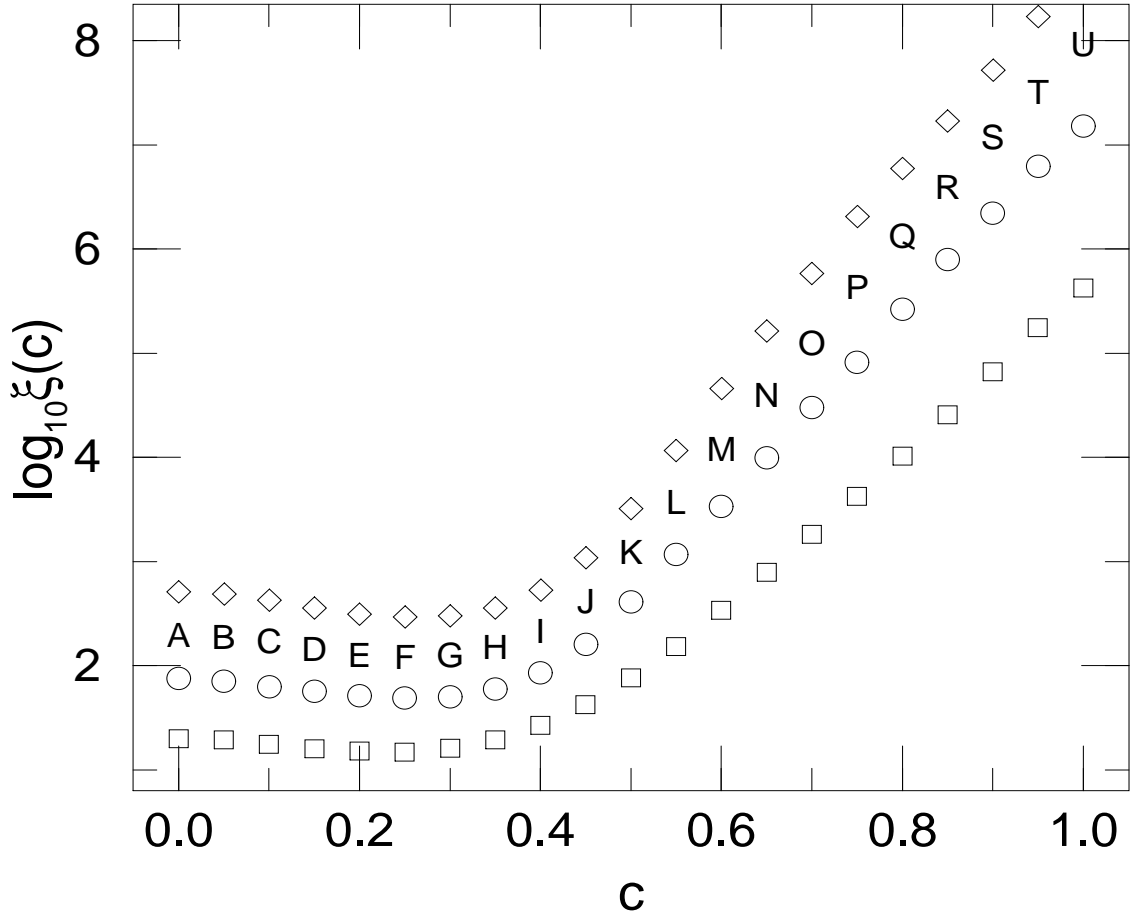,width=7.0in,height=5.5in}}
  \caption{
    Scaling parameter $\xi$ as a function of off-diagonal disorder
    center $c$ at $E= 0.0$ and diagonal disorder $W= 0.0001$
    ($\diamond$), $0.001$ (characters), $0.01$ ($\circ$) and $0.1$
    ($\Box$). }
\label{fig-sca-diag}
\end{figure}

\begin{figure}
  \centerline{\psfig{figure=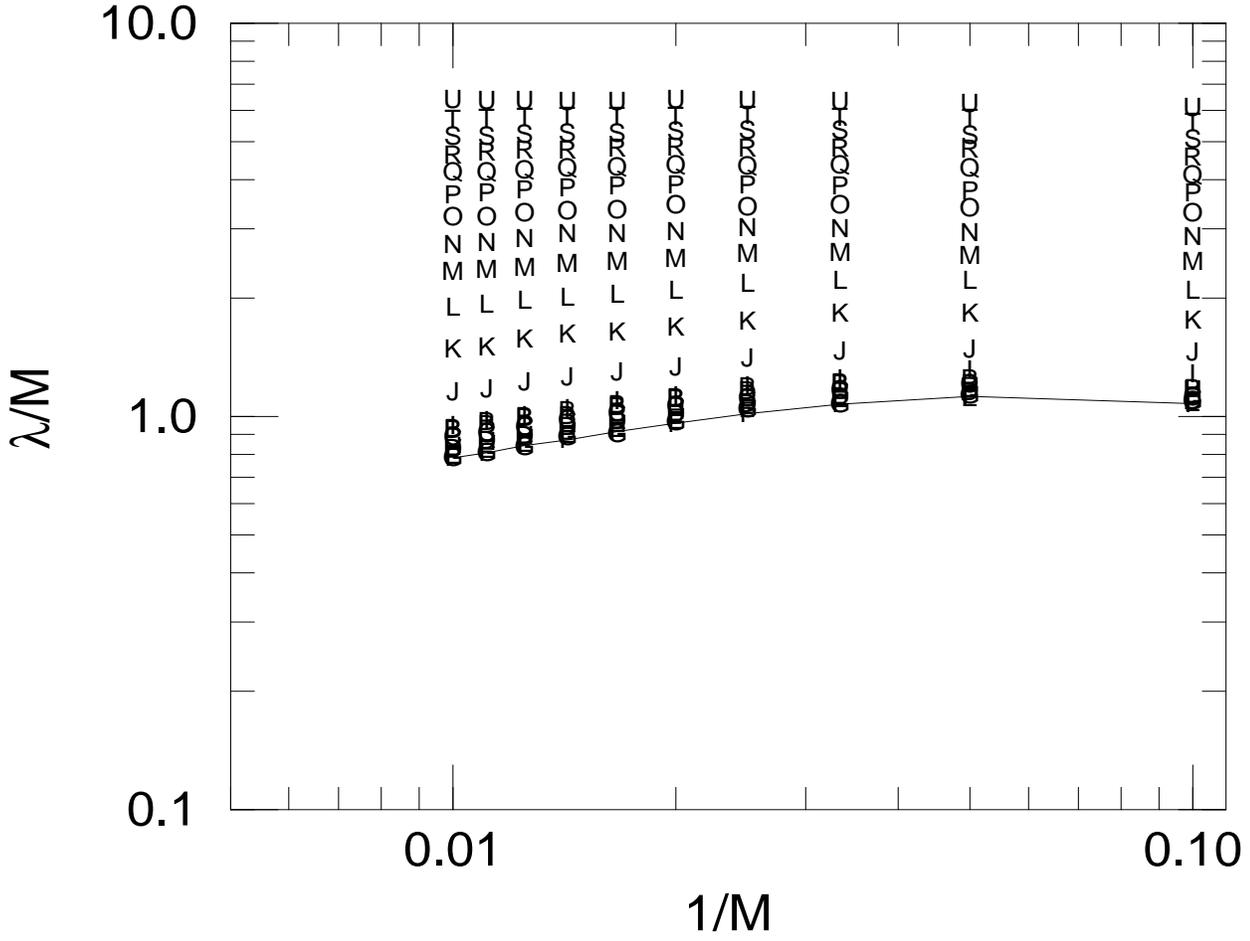,width=7.0in,height=5.5in}}
  \caption{
    Reduced localization lengths $\lambda(M)/M$ vs.\ $1/M$ at $E=0$
    for random hopping and an additional small potential disorder $W=
    0.0001$. The $\lambda/M$ scale is the same as in Fig.\ 
    \protect\ref{fig-fss-diag1} for easier comparison. The line
    connects data corresponding to $c = 0.25$.}
\label{fig-fss-diag2}
\end{figure}

Since it is known that all states are localized in the 2D Anderson
model with purely diagonal disorder --- albeit with fairly large
localization lengths \cite{pichard,mckinnon} --- it is natural to ask
whether the critical state identified above for $E=0$ and purely
off-diagonal disorder is stable against a small additional diagonal
disorder. We thus also performed TMM calculations in which a small
amount of diagonal disorder was used in addition to the off-diagonal
disorder with $w=1$. In Fig.\ \ref{fig-fss-diag1} we show FSS curves
obtained for various small diagonal disorder strengths $W \neq 0$ {\em
  in the band center} $E=0$. Just as for $E\neq 0$ and $W=0$, there is
very nice FSS showing a single scaling curve corresponding to
localization. We note that the values of the scaling parameter $\xi$
for the diagonal disorder $W=0.001$ as shown in Fig.\ 
\ref{fig-sca-diag} are about 2 orders of magnitude larger than for a
2D Anderson model with purely diagonal disorder
\cite{pichard,mckinnon}. This explains why we needed at least an order
of magnitude more transfer-matrix multiplications in our present study
than for purely diagonal disorder.
For $W=0.0001$, we observe deviations from the FSS curve for all $c$
values with $M < 40$ and thus only show data with $M\geq 40$ in Fig.\ 
\ref{fig-fss-diag1}. Nevertheless, using these data we can still
obtain reasonable values for the scaling parameter as shown in Fig.\ 
\ref{fig-sca-diag}.  Also, looking at the values of $\lambda/M$ in
Fig.\ \ref{fig-fss-diag2}, one can see that the reduced localization
lengths decrease as $M$ becomes large again indicating localization.
Only the data with $c=1$ (U) do not yet bend down for larger $M$
values, but rather remain constant and no useful scaling parameter can
be computed.  However, we expect a decrease of $\lambda/M$ for even
larger values of $M$. Thus we are led to the conclusion that even a
very small amount of additional diagonal disorder localizes the
critical state at $E=0$.

In the introduction, we had commented on some apparent similarites of
the present random hopping model with the RFM. Indeed, our results are
somewhat similar to the results obtained recently in Ref.\ 
\cite{batsch} by exact diagonalization and subsequent analysis of the
level statistics. However, in Ref.\ \cite{batsch} states remain
critical in a finite energy range around the band center. Furthermore,
the criticality is not immediately destroyed by an additional diagonal
disorder, but requires a finite amount $W>0$.

%
%

\section{Conclusions}
\label{sec-concl}

We have studied the 2D Anderson Hamiltonian with off-diagonal disorder
by means of exact diagonalization and the TMM. We find from
participation numbers, multifractal exponents and the localization
lengths that for a box distribution $[ c - w/2, c + w/2 ]$ of the
transfer integrals, the strongest disorder effects exist for
$c/w\approx 0.25$. Differences in the localization properties as
compared to the case of purely diagonal disorder are only quantitative
for energies off the band center and all states remain localized.
However, for the states closest to $E=0$, participation numbers and
multifractal properties show substantial differences, and, when taking
into account the proper scale dependence of the participation numbers,
both methods indicate the existence of {\em critical states} at $E=0$
up to the 2D system size $200 \times 200$ for the off-diagonal
disorder.
A TMM study of quasi-1D strips together with FSS further supports the
existence of this critical behavior up to strip width $M=180$ at $1\%$
accuracy. However, even a very small amount of diagonal disorder is
shown to destroy the criticality. Thus it will most likely not play
any role for the transport properties of materials for which the
Hamiltonian (\ref{eq-hamil}) provides a useful model description. We
also do not find any extended states and thus no MIT. Our study is
thus far restricted to a box distribution for the hopping and
potential disorder elements.  However, we believe that similar results
hold for other distributions and combinations thereof.

\acknowledgements

We are grateful to M.\ Batsch and B.\ Kramer for drawing our attention
to the RFM and subsequent discussions. This work has been supported by
the Deutsche Forschungsgemeinschaft (SFB 393). A.E.\ expresses his
gratitude to the Foundation for Polish Science for a fellowship.

%
%

%
%


%
%

\end{document}